\documentclass{edm_template}
\usepackage[english]{babel}
\usepackage{amsmath}
\usepackage[font=footnotesize,labelfont=bf]{caption}
\usepackage{graphicx}
\usepackage[ansinew]{inputenc} 
\usepackage{xfrac}
\usepackage{amsfonts}
\usepackage{wrapfig}
\usepackage{float}
\usepackage{lmodern}
\usepackage{bbm}
\usepackage{amssymb}
\usepackage{underscore}
\usepackage{tikz}
\usepackage{xcolor}
\usepackage[margin=1cm]{caption}
\usepackage{caption}
\usepackage{url}

\usepackage{hyperref}
\hypersetup{
    colorlinks,
    linkcolor={black},
    citecolor={black},
    urlcolor={black}
}
\usepackage{array}
\usepackage{cancel}
\DeclareMathOperator*{\argmax}{arg\,max}
\usepackage{graphicx}
\usepackage{gauss}
\usepackage{lastpage}
\usepackage{epstopdf}
\usepackage{fancyhdr}
\usepackage{standalone}
\usepackage{grffile}

\usepackage{standalone}

\begin{document}

\title{Sequence Modelling For Analysing Student Interaction with Educational Systems}
%\titlenote{(Does NOT produce the permission block, copyright information nor page numbering). For use with ACM\_PROC\_ARTICLE-SP.CLS. Supported by ACM.}}
%\subtitle{[Extended Abstract]
%\titlenote{A full version of this paper is available as
%\textit{Author's Guide to Preparing ACM SIG Proceedings Using
%\LaTeX$2_\epsilon$\ and BibTeX} at

\numberofauthors{5} %  in this sample file, there are a *total*
\author{
% 1st. author
Christian Hansen, Casper Hansen, Niklas Hjuler, Stephen Alstrup, Christina Lioma\\
%\titlenote{Dr.~Trovato insisted his name be first.}\\
       \affaddr{Department of Computer Science}\\
			 \affaddr{University of Copenhagen, Denmark}\\
       %\affaddr{Wallamaloo, New Zealand}\\
       \email{\{chrh,bnq,hjuler,s.alstrup,c.lioma\}@di.ku.dk}       
% 3rd. author
%\and  % use '\and' if you need 'another row' of author names
% 4th. author
%\alignauthor Lawrence P. Leipuner\\
       %\affaddr{Brookhaven Laboratories}\\
       %\affaddr{Brookhaven National Lab}\\
       %\affaddr{P.O. Box 5000}\\
       %\email{lleipuner@researchlabs.org}
% 5th. author
%\alignauthor Sean Fogarty\\
       %\affaddr{NASA Ames Research Center}\\
       %\affaddr{Moffett Field}\\
       %\affaddr{California 94035}\\
       %\email{fogartys@amesres.org}
% 6th. author
%\alignauthor Charles Palmer\\
       %\affaddr{Palmer Research Laboratories}\\
       %\affaddr{8600 Datapoint Drive}\\
       %\affaddr{San Antonio, Texas 78229}\\
}

\maketitle

\begin{abstract}
The analysis of log data generated by online educational systems is an important task for improving the systems, and furthering our knowledge of how students learn. This paper uses previously unseen log data from Edulab, the largest provider of digital learning for mathematics in Denmark, to analyse the sessions of its users, where 1.08 million student sessions are extracted from a subset of their data.
We propose to model students as a distribution of different underlying student behaviours, where the sequence of actions from each session belongs to an underlying student behaviour. We model student behaviour as Markov chains, such that a student is modelled as a distribution of Markov chains, which are estimated using a modified k-means clustering algorithm.
The resulting Markov chains are readily interpretable, and in a qualitative analysis around 125,000 student sessions are identified as exhibiting unproductive student behaviour. Based on our results this student representation is promising, especially for educational systems offering many different learning usages, and offers an alternative to common approaches like modelling student behaviour as a single Markov chain often done in the literature.
\end{abstract}

%% A category with the (minimum) three required fields
%%A category including the fourth, optional field follows...
%\category{D.2.8}{Software Engineering}{Metrics}[complexity measures, performance measures]

%\keywords{ACM proceedings, \LaTeX, text tagging} % NOT required for Proceedings
\section{Introduction and related work}
How students interact with educational systems is today an important topic. Knowledge of how students interact with a given system can give insight in how students learn, and directions for the further development of the system based on actual use. The interaction can be studied both by explicit studies \cite{HuttMWDD16} directly observing student interaction \textit{in situ}, or by the use of log data collected automatically by the use of the system as done in this paper. 
\\ \\
Analysis of log data is often viewed as an unsupervised clustering problem at the student level \cite{FauconKD16, Klingler16}. Our work takes another direction and focuses on the action sequence level.
%[takes some from here http://www.fim.uni-linz.ac.at/staff/paramythis/papers/user655_preprint.pdf]. 
For clustering sequences, Markov models are popular as they provide a convenient way of modelling the transitions and dependencies of the sequences \cite{kock2011}.
%[http://www.fim.uni-linz.ac.at/staff/paramythis/papers/user655_preprint.pdf, top of 2.3]. 
For action sequence mining, both hidden and explicit models have been used depending on the tested hypothesis, and on whether the states are explicit or implicit. Beal et al. use hidden Markov models for student prediction, assuming underlying hidden states of engagement, which can be clustered \cite{beal2007}. K{\"o}ck and Paramythis use explicit states for analysing problem solving activity sequences, as the states in this case are explicit and therefore appear directly in the log \cite{kock2011}.
\\ \\
The choice of clustering of the Markov models depends on the application area. Klingler et al. did student modelling by the use of explicit Markov chains, and the clustering was done by different similarity measures defined on the Markov chains themselves \cite{Klingler16}, e.g. euclidean distance between transitional probabilities, or Jensen-Shannon Divergence between the stationary probabilities of the chains. When individual sequences are clustered, an underlying assumption of the data coming from a mixture of Markov chains has been used \cite{yang2005}, where the individual chains represent the cluster centres, and the task is finding both the chains and the mixing coefficients. 
\\ \\
The work presented in this paper is using discrete Markov chain models for action sequence analysis, on log data\footnote{The data is proprietary and not publicly available} acquired from the company Edulab. Edulab is the largest provider of digital learning for mathematics in Denmark, having 75\% of all schools as customers, and receiving more than 1 million student answers a day. Using a mixture of Markov chains, we assume that each chain will represent a prototype student behaviour. So the underlying assumption in this work is that each student can be modelled as behaving according to some underlying behaviour during each session, and a student can then be seen as a distribution over different behaviours. Edulab's product offers many different ways of learning mathematics, ranging from question-heavy workloads to video and text lessons, and other activities depending on whether the student is in class or at home. This allows to model a student as "distributed" over different behaviours, in contrast to a single student behaviour model of how the student usually interacts with the system.

We reason that mixture of Markov chains will allow for a qualitative study of what type of behaviour each chain represents, and thus ultimately it can be used to show how a student uses the educational system. 
%and a grouping of the students based on their respectively distributions. (HERE NOTE IF WE GET GOOD SEPERATION!!!)
\\ \\
Mixtures of Markov models can be solved by the EM algorithm, which however is notoriously slow to run for large amounts of data, and only local optimal solutions are found \cite{gupta2016mixtures}. In this paper we need fast processing in order to analyse the large amounts of data produced by Edulab, so we simplify the assumptions on the underlying Markov chains, which allows for a modified version of k-means clustering.

Initial cluster centres, representing underlying student behaviour, can be chosen by domain experts and then refined through the clustering. However, since the true number of underlying clusters is unknown, it is difficult for an expert to predefine sensible cluster centres for a range of different numbers of clusters. In this work we first perform simulations to consider the effect of starting at the correct locations versus adding noise to the correct location until the starting points are completely random. Based on these results clustering is done on the Edulab dataset, and a qualitative analysis is performed on the resulting Markov chains. This shows how students are distributed among the Markov chains, and how unproductive system usage can be detected using the Markov chains.

In summary the primary research questions this paper addresses are: 1) to what extent can students be modelled as a distribution over underlying usage behaviours which is changing across sessions, and 2) how this modelling leads to insight in future improvements of the system for the producers of educational systems.
\section{Data}
The data used in this work is produced by matematikfessor.dk, a Danish mathematics portal made by Edulab that spans the curriculum for students aged 6 to 16. The website offers both video and text lessons in combination with exercises covering the whole curriculum, such that it can be used as a primary tool for learning, and not only supplementary. Log data generated by the grade levels corresponding to students of age 12 to 14 for the 2016 school year is used (from August 2016 to February 2017). An action in this system can either be watching a lesson, which contains either a video or text description, or answering a question. Lessons and questions both have a topic id, specifying the general topic of the question or lesson. The data statistics are summarized in Table \ref{dataTable}. The lessons and questions can be assigned as homework or done freely by the students (this study does not differentiate between whether it is homework or not). It should be noted that a lesson takes significantly longer time doing than answering a question hence the lower ratio of lessons, compared to other actions, in Table \ref{dataTable}. %The questions being answered originates from a variety of products, common for the products is that they all present questions to the students, but this study does not differentiate between the products. 
%Future work will look into leverage this additional information.

\begin{figure}
  \begin{center}
    \includegraphics[width=0.5\textwidth]{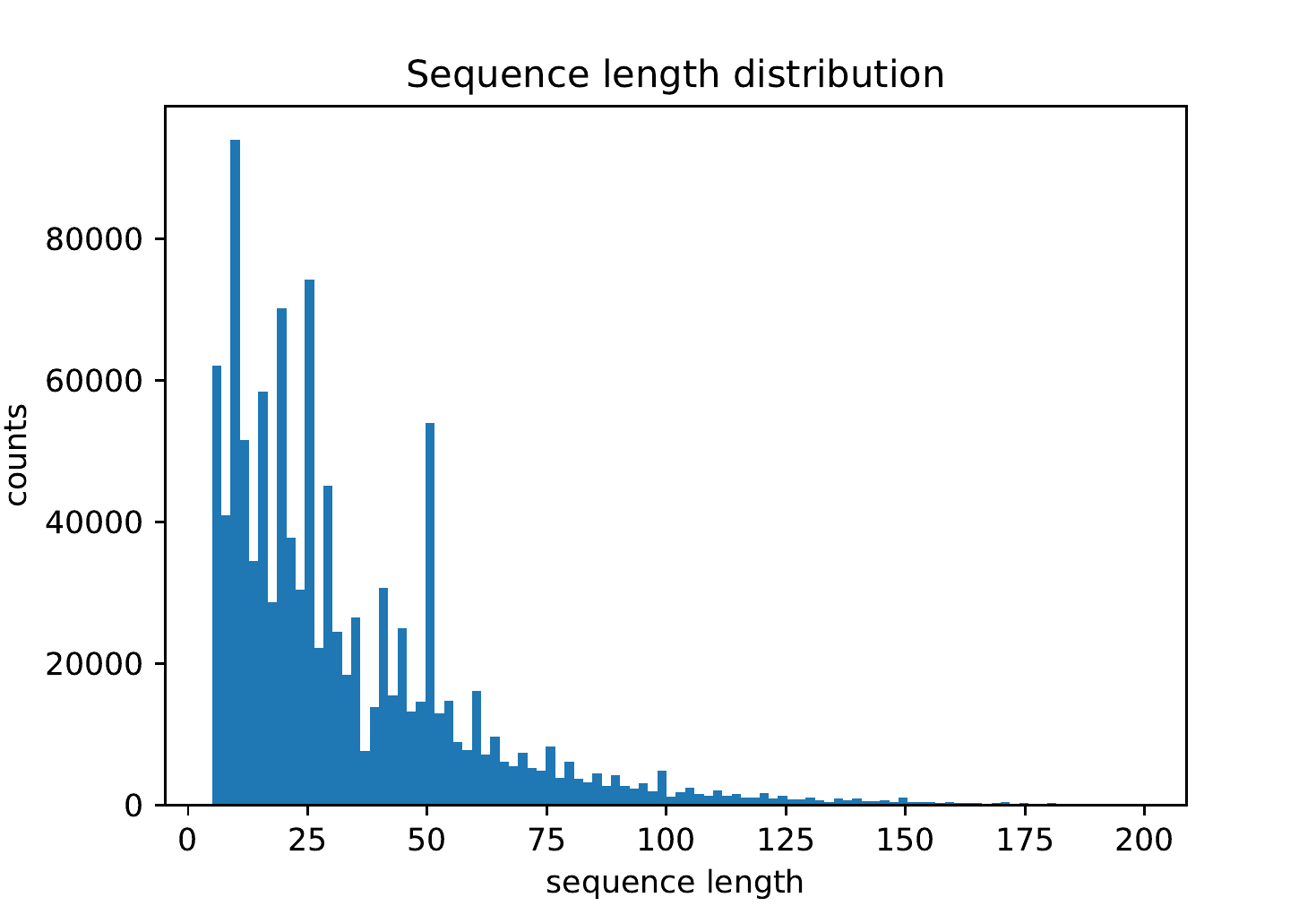}
    \caption{\label{slen}The distribution of action sequence lengths with lengths larger than 200 removed.}
  \end{center}
\end{figure}

The logs do not contain information about when a session is started or finished, so we define a session as a sequence of actions, where the time between two actions is less than 15 minutes. A student has on average 12.5 sessions (standard deviation of 13.3), and the histogram of the number of actions in each action sequence can be seen in Figure \ref{slen}, where sequence lengths larger than 200 have been removed from the plot for the purpose of visualization.
\begin{table}
\begin{center}
\begin{tabular}{|l|c|}
\hline
Number of sequences & 1.08M \\ \hline
Number of actions & 37.5M \\ \hline
Number of lessons & 1.35M \\ \hline
Number of correctly answered questions & 27.44M \\ \hline
Number of wrongly answered questions & 8.71M \\ \hline
\end{tabular}
\caption{\label{dataTable} Data statistics. The number of lessons and question answers sum to the number of actions.}
\end{center}
\end{table}
When a student interacts with the system his actions are stored and seen as an action sequence, an example of one is: 
\begin{align}
\label{actionseq_simple}
Qr_1^{t_1},Qw_2^{t_2},L_3^{t_1},Qw_4^{t_3}, Qr_5^{t_1}, Qr_6^{t_1},Qr_7^{t_1} 
\end{align} 
\(Qr\) is a correctly answered question, \(Qw\) is an incorrectly answered question, and \(L\) is a lesson. The subscript denotes the action number in a temporal ordering, and the superscript denotes the topic id, which is associated with each lesson and question.

\section{Method}
\begin{figure}
  \begin{center}
    \includegraphics[width=0.5\textwidth]{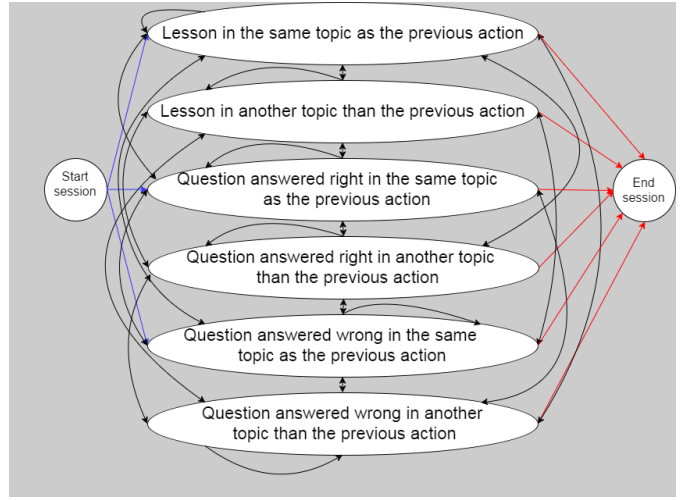}
    \caption{\label{fig:Graph}Markov chain representing the possible states and transitions. Note the transitions each way do not have to be equal.}
  \end{center}
\end{figure}

Our method for action sequence clustering will be explained in this section, and is based on modelling interactions with the system as Markov chains. Our Markov chain model with its transitions is shown in Figure \ref{fig:Graph}. Our model consists of 8 states as will now be explained with their abbreviations in parentheses. These abbreviations are used for visualizing the resulting Markov chains from the clustering. The first two are start (S) and end (E). The rest consists of three general states: Doing a lesson (L), answering a question right (Qr), or answering a question wrong (Qw). Each lesson and question have an associated topic id, which might change from action to action creating the last three states: doing a lesson in another topic than the previous action (L_c), answering a question right in another topic (Qr_c), and answering a question wrong in another topic (Qw_c). If we consider the sequence described in Equation \ref{actionseq_simple}, then that would correspond to visiting the following states
\begin{align}
S \rightarrow 
Qr \rightarrow 
Qw\_c \rightarrow
L\_c \rightarrow \nonumber
\\ 
Qw\_c \rightarrow
Qr\_c \rightarrow
Qr \rightarrow
Qr \rightarrow
E
\end{align}
The pipeline for clustering has the following procedure.
\begin{enumerate}
    \item For every session we extract a sequence of actions $A_1,...,A_n$, and each action sequence corresponds to a path in the used Markov chain model.
    \item Since the Markov chains are unknown, priors $P_1,...,P_k$ (which themselves are Markov chains) are generated at random such that each edge shown in Figure \ref{fig:Graph} has a transition probability taken uniformly at random from 0 and 1. Each random chain is normalized such that each state's outgoing transitional probabilities sum to one. These priors function is the pendant to the usual initial cluster centers, which most often are random data points. Generating a Markov chain from a randomly chosen point would however not work in our case, since many zero valued transition probabilities would occur.
    \item Each action sequence is assigned to the prior which was most likely to generate it, i.e. 
	\begin{align}    
    \argmax_{ 1\leq j\leq k} \left(\prod_{i=1}^m p_{b_{i-1},b_{i}}^j \right)
    \end{align}
    where $p_{b_{i-1},b_{i}}^j$ is the transition probability from state $b_{i-1}$ to $b_i$ in prior $P_j$, $m$ is the number of transitions between states, and $k$ is the number of priors.
    \item After each action sequence has been associated with a prior, then each prior is updated by generating the Markov chain most probable given its associated action sequences. This is done by counting the state transitions in each sequence in a new Markov chain model, and normalizing afterwards.
   % $e_{a,b}^j = \frac{\text{\# transitions from state a to state b for all paths in cluster } C_j}{\text{\# transitions from state to a to any other state for all paths in cluster } C_j} $
    \item Points 3 and 4 are ideally reiterated until convergence, i.e. no action sequence changes its associated prior. However for computational reasons we stop iterating after less than 5\% of the sequences have changed their assigned prior.
\end{enumerate}
The clustering technique is very similar to ordinary k-means clustering, with the major difference that the clustering is not dependent on a similarity measure directly on the sequence, but dependent on the Markov chains generated by the clustering. Comparing to ordinary k-means clustering, the produced chains in each iteration are analogous to the ordinary cluster center found by some mean. The mixture model could also be estimated by the EM algorithm \cite{Barber2012}, which has the benefit that sequences that do not belong to a single clear cluster, i.e. that have multiple highly probable chains, will weight in on all of them. This has the downside that clusters take longer to be separated, and the convergence is therefore slower. Under the assumption of the chains being distinct, each sequence will mostly weight on a single chain, and here the k-means clustering method and EM algorithm will perform very similarly. For the data from Edulab we assume most of the chains to be distinct, but not necessarily all. In addition a very large number of sequences will have to be clustered in the future when the full dataset is used, and not restricted as done for this paper. We are therefore mostly interested in how well the k-means clustering approach performs as it is more computationally feasible when the data size is increased.

The above procedure leaves two challenges: 1) How do we know the resulting Markov chains are close to the real ones? and 2) How to estimate the number of priors? We address these points next.

The first point is dealt with using synthetic data, where k random Markov chains are made, and each action sequence is generated from one of those chosen uniformly at random. In order to ensure a suitable length of the generated action sequences, the ingoing probabilities to the end state are fixed to allow for an average sequence length of 20. After generating the synthetic data, the most probable Markov chain for each sequence is assigned as its label, and the goal in the clustering is to be able to capture these clusters. Note, that since each sequence is randomly generated using the chosen Markov chain, then its most probable Markov chain might not be the one generating it. 
To determine the ability to capture the original clusters we consider the average purity of the resulting clusters:
\begin{align}
    Average_{purity} =\frac{1}{n} \sum_{i=1}^n \frac{\max_{1 \leq j\leq k}(|C_j \cap S_i|) }{|S_i|} 
\end{align}
Where $S_i$ is an estimated cluster, $C_j$ is the true cluster, $n$ is the number of clusters, and $k$ is the number of true clusters. An average purity of 1 represents that the method fully captures the original clusters. The underlying Markov chains are unknown on real data, so increasingly noisy versions of the underlying Markov chains are experimented with as priors, to show how the method is expected to perform under real circumstances.

In the case of real data, the true underlying Markov chains are unknown, so in this case the sum of the log likelihoods is calculated for the sequences to their most probable prior:
\begin{align}
    \text{sum of log likelihood} =\sum_{i=1}^n \log \left( \mathcal{L}(s_i | P_i^*)\right)
\end{align}
where $s_i$ is an action sequence, $P_i^*$ is the prior most likely to generate action sequence $s_i$, and $\mathcal{L}(s_i | P_i^*)$ is the likelihood that $P_i^*$ generates $s_i$. 

The second point mentioned earlier, about estimating the number of priors, can be solved using either the average purity in the synthetic case, or from the sum of log likelihoods in the real case. The sum of log likelihoods as a function of $k$ will be monotonically increasing, but the slope will decrease as $k$ exceeds its true underlying value. Since the method starts with randomly chosen priors, it is repeated a number of times, and the solution with the largest log likelihood is chosen for each value of $k$.

\section{Simulated experiment with \\noisy priors}
There are two approaches for estimating the Markov chains for the Edulab data set. 1) The prior Markov chains can be chosen by domain experts - by specifying common sequences we would expect to find in the data, and then refine them during the clustering. 2) The second approach is as described in the method section, starting with random chains, and running k-means multiple times, and taking the clustering which gives the highest sum of log likelihoods. To measure how the method behaves as the initial priors are increasingly noisy versions of the underlying Markov chains, k-means is run with the priors chosen as:
\begin{align}
P_i = (1-\alpha) P_i^*  + \alpha P_{rand}
\end{align}
Where all \(P\)s are Markov chains represented by matrices of transitional probabilities, and $\alpha$ is the noise parameter. \(P_i\) is the \(i^{th}\) prior, \(P_i^*\) is the \(i^{th}\) underlying Markov chain used when generating the synthetic data, and \(P_{rand}\) is a random Markov chain. The higher \(\alpha\), the more noisy the initial prior is. 

\begin{figure}
  \begin{center}
    \includegraphics[width=0.4\textwidth]{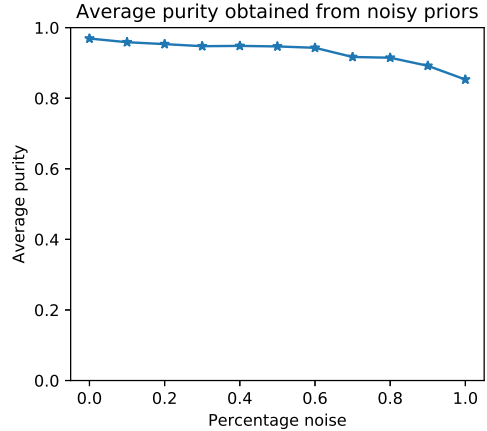}
    \caption{\label{avgpurity} Average purity as a function of increasingly more noisy priors. A completely random prior (1.0 on the x axis) is able to perform well.}
  \end{center}
\end{figure}

In Figure \ref{avgpurity}, we see how the average purity behaves as a function of noise parameter \(\alpha\). The experiment is run for $k=6$, and 6 random chains are generated. The transition probabilities to the end state are fixed at 0.05 for all states for all chains to allow for sequences of average length 20. 50000 sequences are sampled uniformly from the 6 chains. The modified k-means is then run with the priors varying depending on \(\alpha\), and the experiments are run 10 times and purity is the average over the 10 runs.
First we note that even with using the modified k-means algorithm and not the EM algorithm the resulting average purities are quite high. It is seen that even with \(\alpha=1\) representing completely random priors, the reduction in purity is not too large compared to starting with the same priors as the data is generated from. Even starting with the same priors which generated the data does not guarantee perfect purity, which is expected as there are some sequences that are almost as likely under multiple chains, so small differences in the data determined Markov chains will move them from one chain to another. Based on the above result we will not define the priors by an expert, and instead let them be random. This has the benefit of being more manageable than hand-crafting specific priors for each choice of k, which would be very difficult to do in a meaningful way when k is large.

\section{Real data experiment}
\subsection{Choosing the number of clusters}
The problem of determining the number of clusters is common for all unsupervised learning tasks. In this paper we consider the sum of the log likelihoods for the action sequences. A common approach is the use of the "elbow" heuristic, where the choice of k is chosen based on the slope of the sum of log likelihoods function over k. %The elbow heuristic is often used, as it does not require explicit calculations of the complexity class of the clustering over k, which information criterion approaches would need. 

In order to argue that there is structure in the data, and that the method is able to capture this structure, a randomized experiment is made. The randomized experiment consists of randomly permuting each sequence (but keeping the start and end states), and seeing how the sum of log likelihoods is affected by it. If there is no structure originally in the sequences, then one can not expect it to perform better than the permuted data.
\begin{figure}
  \begin{center}
    \includegraphics[width=0.47\textwidth]{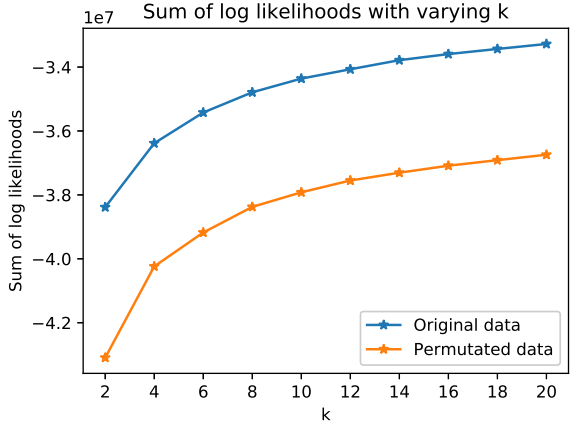}
    \caption{\label{sumloglike} Sum of likelihoods for the best performing clusters for each k. Each experiment is run 5 times for each k. The permutations of each sequence is done for each value of k in each of the 5 times. }
  \end{center}
\end{figure}

In Figure \ref{sumloglike} we see that the sums of log likelihoods are considerably lower in the permuted data set, with only slightly higher sum of log likelihoods when $k=20$ compared to $k=2$ for the real data set. The action sequences therefore have structure which the Markov chain captures, and it is therefore not just random chains that the k-means clustering produces. Since the chains capture some inherent structure in the data, it is meaningful to analyse the individual chains with regards to what user behaviour they capture.

There is not an obvious breaking point in the sum of log likelihoods, but the increase before $k=6$ is large, while the increase for $k>10$ is notably smaller, so a value of k between 6-10 is sensible. We will in the qualitative assessment of the chains use $k=6$.
%since a lower number of chains are easier as the qualitative assessment require inspection and comparisons of the chains. Ovejvej bedre forklaring
\subsection{Qualitative assessment of Markov chains}
\begin{figure}
  \begin{center}
    \includegraphics[width=0.48\textwidth]{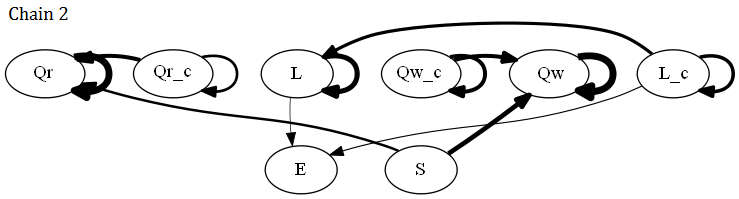} %1
    \includegraphics[width=0.48\textwidth]{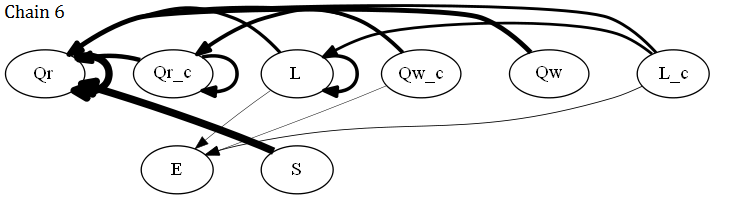} %5
    \includegraphics[width=0.48\textwidth]{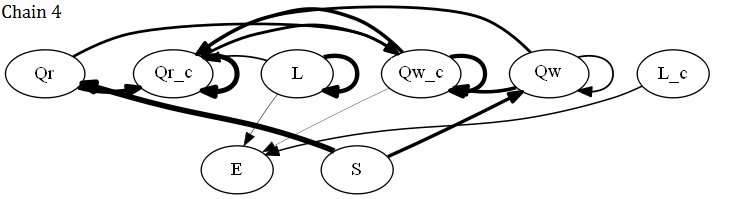} %3
    \caption{\label{chains}Chains 2, 6, and 4 of the six chains. The thickness of the arrows is proportional to the transitional probability for each state, except the ending state. The transitional probabilities are sorted and only drawn until 70\% of the probability mass is covered. For the ending state 70\% of the incoming transitional probabilities are drawn. State abbreviations are explained in section 3.
%The first chain indicates very little mixing between states, second chain indicates usage where most actions leads to correct answered questions and the user primarily focuses on 1 topic, third chain indicate more mixing between topics and between being correct and wrong. The first chain is based on 126683 sequences, the second chain is based on 194174 chains, and the last chain is based on 131460 sequences.
    }
  \end{center}
\end{figure}

This section will make qualitative assessments of what the different resulting Markov chains represent with regards to what type of user behaviour they capture. Even with six chains there is some similarity between some chains, so in this section we will focus on the three most distinct chains shown in Figure \ref{chains}. The thickness of the arrows is proportional to the transitional probability for each state, except the ending state. The transitional probabilities are sorted and only drawn until 70\% of the probability mass is covered. For the ending state, 70\% of the incoming transitional probabilities are drawn.

In general not all chains can be described as either being a positive or negative usage of the system. Chain 2 captures usage where most of the questions being answered are either right or wrong, and there is very little mixing between taking lessons and answering a question. Usage like this could indicate an unproductive session for students, since they are mostly getting all questions right or all questions wrong, and research shows that students feel more intrinsic pleasure when the difficulty level is slightly challenging \cite{gott2013} leading to more engaged sessions \cite{csik1992}. Similarly, watching lessons without engaging with the material via questions leads to students not training the learned material, which is important for the learning process.

Chain 6 can be described as a positive usage of the system, as the most probable transitions lead to a question being correctly answered, except for the two transitions in the lessons. Generally students are focused on one topic at a time. 

Chain 4 has high transitional probability when switching between topics, so this could indicate a session with a primary focus on repetition as the topic is varying, and students most often answer questions from another topic than the watched lessons.
\\ \\
The distribution of the sessions over the chains can be seen in Table \ref{distOverChains}.	

%\begin{table}
%\begin{center}
%\begin{tabular}{|l|c|c|c|c|c|c|}
%\hline
%Chain number & 1 & 2 & 3 \\ \hline
%Num. sequences & 295,792&   126,683&   198,736 \\ \hline
%Avg. sequence length & 34.81&  36.88&  26.79 \\ \hline  \hline
%Chain number & 4 & 5 & 6 \\ \hline
%Num. sequences &  131,460&   194,174&  144,121 \\ \hline
%Avg. sequence length &  28.79&   36.12&  44.85 \\ \hline
%\end{tabular}
%\caption{\label{distOverChains} The number of sequences and average length of sequences for each Markov chain}
%\end{center}
%\end{table}

\begin{table}
\begin{center}
\begin{tabular}{|l|l|l|}
\hline
& Num. sequences & Avg. sequence length \\ \hline
chain 1 & 295,792  & 34.81 \\ \hline
chain 2 & 126,683 & 36.88 \\ \hline
chain 3 & 198,736 & 26.79 \\ \hline
chain 4 & 131,460 &  28.79   \\ \hline
chain 5 & 194,174 & 36.12 \\ \hline
chain 6 & 144,121 &  44.85 \\ \hline
\end{tabular}
\caption{\label{distOverChains} The number of sequences and average length of sequences for each Markov chain}
\end{center}
\end{table}

\begin{figure}
\
  \begin{center}
    \includegraphics[width=0.5\textwidth]{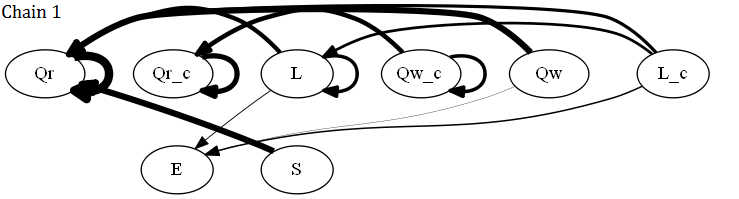} %1
    \caption{\label{mostOcChain} Chain 1, the most common chain. State abbreviations are explained in section 3.
    }
  \end{center}
\end{figure}

The length of the sequences is varying, but no single chain in general captures either the very short or very long sequences. Instead a combination of shorter and longer sequences is captured by each chain. The most common chain can be seen in Fig \ref{mostOcChain}. 
This chain is similar to chain 4 (Fig \ref{chains}), but with more topic changes and more wrongly answered questions when changing topics, which can be seen in the self loop for \(Qw\_c\). Chain 4 is also shorter on average. As seen in Table \ref{distOverChains}, generally all six chains contain a large amount of sequences on average. This indicates that the system usage does indeed vary, and is not limited to all sequences of the same length defining the same use of the system. If one considers each user's distribution of Markov chains, then on average each user has 3.5 different types of sessions out of 6 with a standard deviation of 1.5. This supports the assumption that a single Markov chain is not optimal for user profiling for educational systems similar to the one generating our data, where there is a lot of user freedom in what activities they engage in.
\section{Discussion and Conclusion}
In this work first order Markov chains have been used, but it is generally known that the action sequences do not fulfil the Markov property of transition to a state only being dependent on the previous state. No order of Markov chain will completely capture the underlying transition between states, as the usage is dependent on many external factors which are unknown, but higher order chains would be able to capture more complex dynamics in the usage. Even though the Markov property is violated, Markov chains are still very widely used in educational data mining \cite{FauconKD16, Klingler16}, and provide a good tool for comparisons of action sequences across different lengths, focusing on the flow of actions taken. In future work an interesting extension would be considering time dependent Markov models, such that the transitional probabilities are dependent on how long the states have been unchanging. This would allow for more interpretative models, e.g. we could see when the probability of a session ending gets high.

When inspecting the Markov chains produced by the clustering, chain number 2 indicated suboptimal or unproductive usage of the system, where the students either experience questions that are too easy or too hard, or never train what they learn in the lessons.  The chain has 126,683 sessions in its cluster, and it is therefore a significant amount of sessions where the learning outcome most likely could be improved. Based on this it could be recommended to have a few obligatory questions after a lesson to strongly encourage the student to use what they have just learned, and detect negative spirals where the students are always wrong by recommending lessons to help the student move forward.

Modelling the student as a distribution over Markov chains, which can be considered usage patterns, results in a vector representation of the individual students. This representation allows to apply standard techniques directly on the student model, compared to working on more complex student models. An example is the issue of drift in student behaviour over time, corresponding to some learning, or wider cognititive development of the student. This problem has also been considered in a similar context in \cite{Klingler16}, where distances between single Markov chains on a student level were estimated. However, in our setting standard methods could readily be used to detect this type of drift and potentially alert the teacher. 

The work presented shows a qualitative study of the proposed student representation, and experiments using synthetic data show that our methodology is able to capture the underlying generative Markov chains very well, when the number of chains has been estimated. A source for future work will be using the student vectors in a predictive task, such that quantitative measures can be acquired. An interesting path would be using knowledge tracing methods over the different session types, to see if there are any unexpected differences between the knowledge acquired by the student depending on the type of session - i.e. the kind of Markov chain the session originates from. 

%In general we believe the idea of vectorizing student behaviour over prototype usage patterns instead of traditional measures as correct question percentage and similar is an interesting direction, that can allow for more flexible student models.

%\end{document}  % This is where a 'short' article might terminate

%ACKNOWLEDGMENTS are optional
\section{Acknowledgments}
The work is supported by the Innovation Fund Denmark through the DABAI project. 
\bibliographystyle{abbrv}

\balancecolumns
% That's all folks!
\end{document}